\documentclass[prb,aps,twocolumn]{revtex4}
\usepackage{tabularx}
\usepackage{bm}
\usepackage{euscript}
\usepackage{epsfig,psfrag,subfigure}
\usepackage{graphicx}
\usepackage{color}
\usepackage{amsmath}
\usepackage{amsfonts}
\usepackage{exscale}
\usepackage{amsbsy}
\usepackage{stmaryrd}
\usepackage{wasysym}

\def\be{\begin{equation}}       \def\ee{\end{equation}}
\def\bea{\begin{eqnarray}}      \def\eea{\end{eqnarray}}
\def\ba{\begin{array} }
\def\ea{\end{array} }
\def\bnum{\begin{enumerate} }
\def\enum{\end{enumerate}}

\def\nn{\nonumber}

\def\=>{\Rightarrow}
\def\>{\rightarrow}

\def\eye2{Fathbb{I}}

\def\BNA{Ba$_{1-x}$Sr$_x$Ni$_2$As$_2$}

\newcommand{\eq}[1]{\begin{align}#1\end{align}}

\begin{document}

\title{\bf Tests of nematic-mediated superconductivity applied to \BNA}
\author{S. Lederer$^1$, Erez Berg$^2$, and Eun-Ah Kim$^1$}

\affiliation{$^1$Department of Physics, Cornell University, Ithaca, New York 14850, USA}
\affiliation{$^2$Department of Physics, Weizmann Institute of Science, Rehovot, Israel}
\begin{abstract}
In many unconventional superconductors, nematic quantum fluctuations 
are strongest where the critical temperature is highest, inviting the conjecture that nematicity plays an important role in the pairing mechanism. 
Recently, \BNA\  has been identified as a tunable nematic system that provides an ideal testing ground for this proposition.
We therefore propose several sharp empirical tests, supported by quantitative calculations in a simple model of \BNA. The most stringent predictions concern experiments under uniaxial strain, which has recently emerged as a powerful tuning parameter in the study of correlated materials. Since uniaxial strain so precisely targets nematic fluctuations, such experiments may provide compelling evidence for nematic-mediated pairing, analogous to the isotope effect in conventional superconductors.
\end{abstract}
\date{\today }
\maketitle

 \emph{\bf Introduction}: Considerable debate remains regarding the pairing mechanisms of most unconventional superconductors. Many hypothesize that fluctuations of some order parameter(s), such as magnetism, mediate the bulk of the pairing interaction. These hypotheses are plausible, in part, because the superconducting region of the phase diagram is often close to various other forms of long range order. A ubiquitous form of such order is nematicity, which breaks the discrete point group symmetry of the crystal lattice. Theoretical studies have established, in the abstract, that nematic fluctuations promote superconductivity\cite{Yamase2013,Maier2014,raghu2015,metlitski2015cooper,lederer2015enhancement,kang2016,Wang2016,liu2017,lederer2017,klein2018a,klein2018,Abanov2018,Wu2019}, but their relevance to the superconductivity of any given material is difficult to assess. There is thus a need for distinctive and testable predictions based on the hypothesis of nematic-mediated pairing.
 
 In addition to concrete predictions, testing the hypothesis requires sufficiently simple material systems, preferably ones with tunable nematic fluctuations. For instance, while the Fe-based superconductors exhibit ubiquitous\cite{kuo2016} and tunable nematic fluctuations, these are (with rare exceptions\cite{Abdel-Hafiez2015,Hosoi2016}) accompanied by the confounding factor of a nearby magnetic phase.
 
 Happily, recent experiments have identified a tunable, nonmagnetic model system in the \BNA\ series\cite{ronning2008,sefat2009,Kothapalli2010,eckberg2019}. The nematic susceptibility in the B$_{1g}$ channel\footnote{We here use the crystallographic two-Ni unit cell. Note that the channel of strong nematic fluctuations in iron-based superconductors is $B_{2g}$ in this nomenclature} (measured by elastoresistance) grows as the Sr concentration $x$ is reduced from 1.0 to 0.7, from below the noise floor to a large (dimensionless) value of nearly 20. Meanwhile, the superconducting critical temperature, $T_c$, rises dramatically, from $0.6 K$ to 3.5 $K$. The typical determining factors of $T_c$ in BCS superconductors--the Debye frequency and the density of states--vary negligibly in this range of $x$, leading the authors of [\onlinecite{eckberg2019}] to suggest that nematic fluctuations are responsible for the enhancement. 

 Here, we examine the influence of nematic fluctuations on various superconducting properties, based both on general considerations and on explicit calculations for a simple model of the \BNA \ system, for $0.7<x<1.0$. One set of predictions follows from anisotropy (momentum dependence) of the superconducting gap, which nematic fluctuations promote by virtue of their anisotropic coupling to electrons. The other set concerns the effects of uniaxial strain, which explicitly breaks lattice rotation symmetry and reduces the strength of nematic fluctuations. Both sets of predictions are essentially general, but those regarding strain are sharper, may apply to a broader variety of materials, and are particularly timely in light of the increasing use of strain as an experimental tuning parameter\cite{Hicks2014}.

\begin{figure}%
\centering
\includegraphics[trim = 30 40 20 55, width=7cm,clip=true]{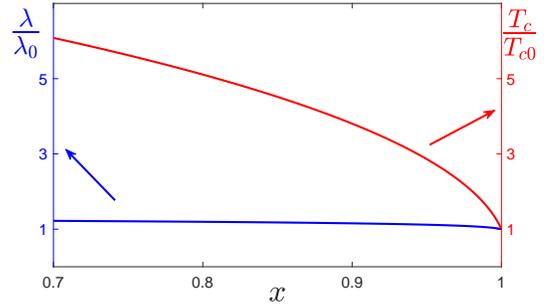}%
\caption{$T_c$ versus doping: The pairing eigenvalue $\lambda$ (in blue) grows only modestly as strontium concentration $x$ is reduced from 1.0 to the critical doping of $0.7$ (where the triclinic phase onsets). By contrast, $T_c$ (in red) grows dramatically.  This difference follows from the essential singularity of $T_c$ in the limit $\lambda\to 0$ (Eq. \ref{eq:Tc}). Here, we assume three dimensional nematic fluctuations, with parameters $\lambda_0=0.1$, $g=4.4$, $k_{1}=\pi/2$, $k_2=\pi/4$, $m_1=m_2=1$, $\chi_0=0.1\xi^2$ and $\Lambda=\pi/6$. Both the extent of $T_c$ enhancement and the shape of the $T_c(x)$ curve are sensitive to the choice of $\lambda_0$, so these results should be understood only qualitatively when compared with experiment.}
\label{fig:Tc}%
\end{figure}

\begin{figure}%
\centering
\includegraphics[trim= 180 0 220 0, clip=true,width=8cm]{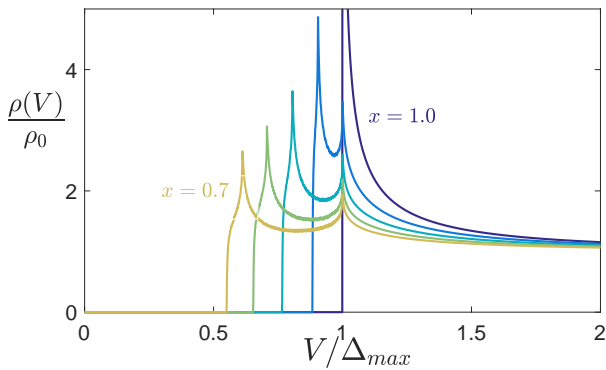}%
\caption{Density of states $\rho$ (approximately proportional to tunneling conductance) vs. energy $V$ at selected values of doping for the parameters of Fig. \ref{fig:Tc}. $\rho$ is measured in units of its normal state value $\rho_0$. Gap anisotropy increases as $x$ is reduced and nematic fluctuations increase. The anisotropy splits the peaks in the density of states, which occur at local maxima of the gap. Slight broadening has been applied for visual clarity.}
\label{fig:dIdV}%
\end{figure}

\begin{figure}%
\centering
\includegraphics[trim=150 10 90 10,clip, width=8cm]{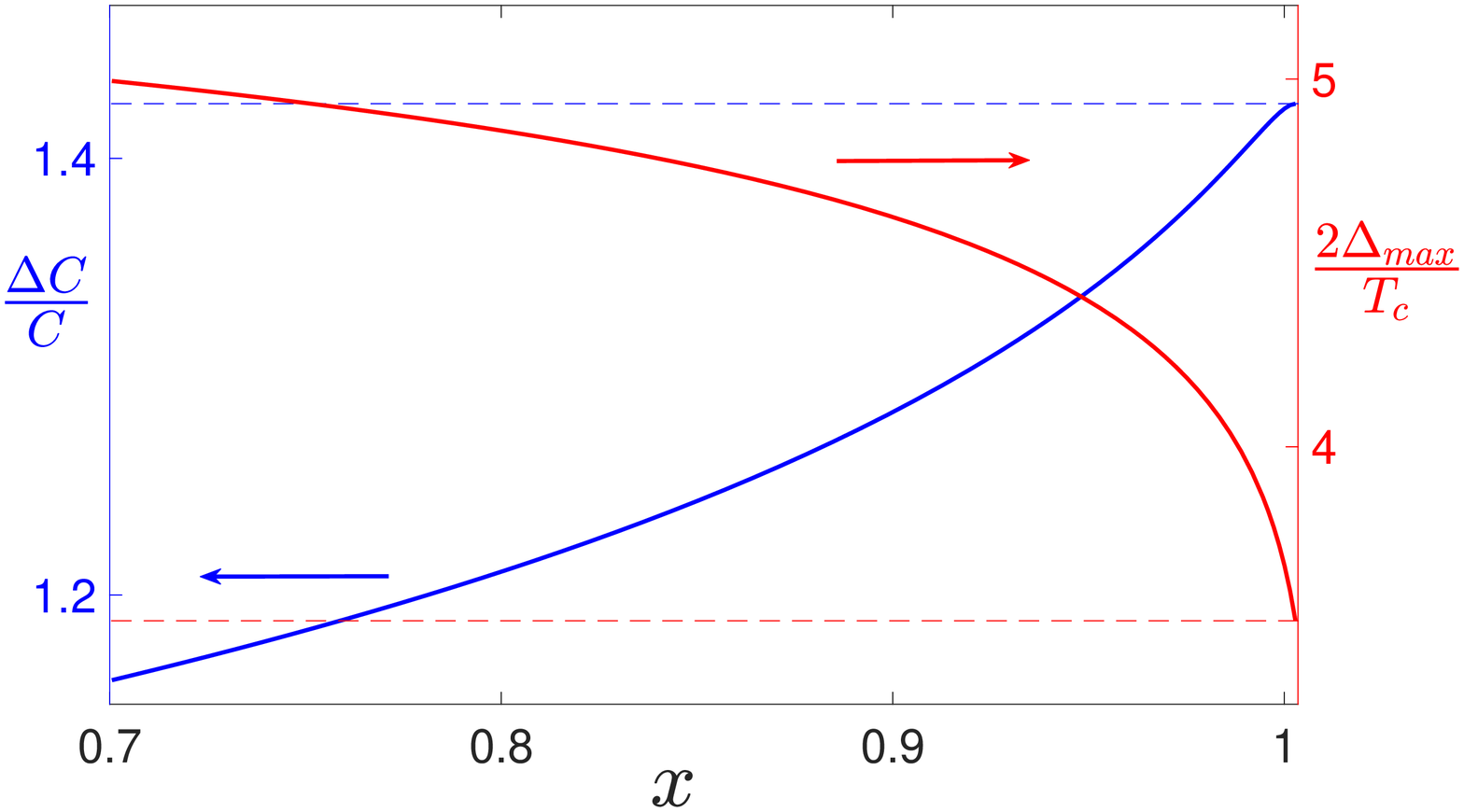}%
\caption{Variation of the fractional specific heat jump and the gap to $T_c$ ratio with doping, for the parameters of Fig. \ref{fig:Tc}. The blue and red dotted lines show the conventional BCS values attained at $x=1.0$. These quantities are computed from formulas derived in the Supplementary Material.}

\label{fig:ratio_jump}%
\end{figure}

\begin{figure}%
\centering
\includegraphics[trim= 30 30 30 30, clip=true,width=7cm]{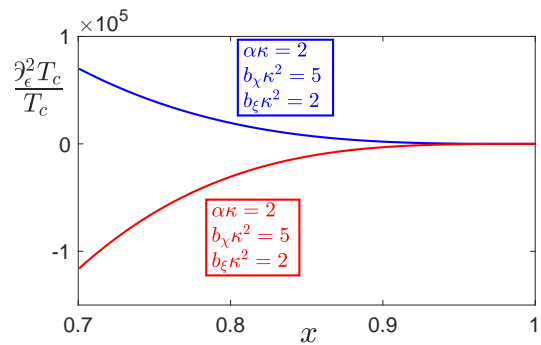}%
\caption{Strain sensitivity versus doping: $T_c$ varies quadratically with applied $B_{1g}$ strain $\epsilon$, but with a coefficient whose magnitude grows vastly with the strength of nematic fluctuations. The strength of the nematic-mediated interaction is proportional to the ratio of susceptibility $\chi_0$ to a power of the correlation length $\xi$ (Eq. \ref{eq:g}). Both of these decrease with strain, so their ratio (and therefore $T_c$) can either increase or decrease, as shown respectively in the blue and red curves above. Parameters are as shown, with others as in Fig. \ref{fig:Tc}.}
\label{fig:strain}%
\end{figure}

\emph{\bf General considerations:}
The influence of nematic fluctuations on superconductivity can be comprehensively analyzed within a weak coupling framework except asymptotically close to a nematic quantum critical point\cite{lederer2015enhancement}. Since there is no indication of a zero temperature divergence of the nematic susceptibility in the $x-T$ phase diagram of \BNA\cite{eckberg2019}, the weak coupling approach may give qualitatively correct results in the entire doping range under consideration. For our purposes, the main results are that nematic fluctuations 1) enhance $T_c$, potentially by a large amount, as shown in Fig. \ref{fig:Tc}; 2) promote a characteristic anisotropy of the superconducting gap, whose consequences are shown in Figs. \ref{fig:dIdV} and \ref{fig:ratio_jump}; and 3) dramatically enhance the influence of uniaxial strain on superconductivity, as shown in Fig. \ref{fig:strain}. In this section, we summarize some key results of the weak coupling approach of [\onlinecite{lederer2015enhancement}], to which we refer the reader for further details.

Consider, for simplicity, a weak coupling s-wave superconductor with density of states $\rho_0$ at the Fermi level. Assume the superconductivity is due to a momentum-independent attractive interaction of strength $V_0$ operating at frequency $\Omega_0$ (this would be the Debye frequency for conventional phonon-mediated attraction). 
Then the gap and $T_c$ are both of order $\Omega_0\exp(-1/\lambda_0)$, where $\lambda_0=\rho_0V_0$ is the pairing eigenvalue.

Now weakly couple the system to a separate nematic mode with a long correlation length $\xi$.  
The nematic fluctuations mediate an attractive interaction $V^{(ind)}_{k,p}$, that is dominated by forward scattering, since the interaction is negligible for momentum transfer greater than $\xi^{-1}$. Accordingly, as $\xi$ increases, $V^{(ind)}_{k,p}$ becomes nearly diagonal in momentum, and can be approximated by a delta function:

\eq{
V^{(ind)}_{k,p}\approx-h(k)\delta(k-p)
}

Here $h(k)\geq 0$ is a coupling function governed by the symmetry of the nematic order parameter, and we will take an explicit form in the next section. $h$ increases as the strength of nematic fluctuations increases. 
Because $ V^{(ind)}_{k,p}$ is diagonal in momentum, it is simple to compute its pairing eigenvalues. The largest is $\lambda^{ind}=\text{max}[h(k)/v_F(k)]$, where $v_F(k)$ is the Fermi velocity and the maximum is computed for $k$ on the Fermi surface. We assume that $\lambda^{ind}\ll\lambda_0$, so that the ``bare" attraction is still the dominant part of the pairing interaction\footnote{The opposite case $\lambda^{ind}\gg\lambda_0$ can also be analyzed straightforwardly. We ignore the effect of retardation, namely that $V^{ind}$ operates at a lower frequency scale than $V_0$. This effect can easily be incorporated, but in weak coupling is parametrically less important than the pairing eigenvalue effects considered here.}.

With these assumptions, the changes in $T_c$ and the gap function can be computed using perturbation theory in $V^{(ind)}$, applied to the linearized gap equation. The new pairing eigenvalue is $\lambda=\lambda_0+\delta\lambda$, where
\eq{
\delta\lambda=\left(\oint\frac{d{\bf k_\parallel}}{v_F^2({\bf k_\parallel})}h({\bf k_\parallel})\right)\left(\oint\frac{d{\bf k_\parallel}}{v_F({\bf k_\parallel})}\right)^{-1},
}
With the integrals over the Fermi surface. The pairing eigenvalue has been increased by a suitably weighted average of the coupling function, but by a small amount $\delta\lambda<\lambda^{(ind)}\ll\lambda_0$. $T_c$ has also been increased:
\eq{
\frac{T_c}{T_{c,0}}=&\exp\left[\frac{1}{\lambda_0}-\frac{1}{\lambda_0+\delta\lambda}\right]\nn\\
\approx &\exp\left[\frac{\delta\lambda}{\lambda_0^2}+{\cal O}\left(\frac{\delta\lambda^2}{\lambda_0^3}\right)\right]
\label{eq:Tc}} 
Note that the enhancement of $T_c$ can be large despite the fact that $\delta \lambda/\lambda_0$ is small, provided $\delta \lambda \gtrsim\lambda_0^2$. The new gap function can also be computed in perturbation theory. To leading order,
\eq{
\frac{\Delta(k)}{\Delta_{max}}\approx 1+\left(\frac{1}{\lambda_0}\right)\left(\frac{h(k)}{v_F(k)}-\left[\frac{h(k)}{v_F(k)}\right]_{max}\right)
\label{eq:wavefunction}}
The gap function is now anisotropic, with the pattern of anisotropy determined by the band structure and $h(k)$, which is in turn governed by the symmetry of the nematic order parameter.

\emph{\bf Model and numerical results}: Here we consider an effective model for the physics of \BNA. In the absence of nematic fluctuations, the model comprises a fermion band structure and a weak attractive interaction, giving rise to conventional BCS superconductivity. We then couple the fermions to a nematic bosonic mode $\phi$ in a symmetry-appropriate way, and specify the ``bare" fluctuation spectrum of $\phi$. Explicit calculations are performed using the perturbative renormalization group approach of [\onlinecite{lederer2015enhancement}].

We model the low energy bandstructure (as measured in angle-resolved photoemission\cite{zhou2011}) with four cylindrical sheets, two centered at the M point and of radius $k_{1}$ and one each at the X and Y points with radius $k_{2}<k_{1}$. We assume parabolic dispersion with masses $m_{1,2}$. Before coupling to nematic fluctuations, we take the interaction between fermions to be a momentum independent attraction of strength $V_0>0$, so that there is BCS superconductivity with $T_c$ and gap proportional to $\exp(-1/\lambda_0)$, with $\lambda_0\propto (m_1+m_2)V_0$ the unperturbed pairing eigenvalue. We couple the fermions to the nematic mode $\phi$ with the term
\eq{
S_{int}=&\alpha\int d\tau dq  dk \bigg[f(k)\phi_q\bar\psi_{k+q/2}\psi_{k-q/2}\bigg],
}
where spin and band indices are implicit. The form factor $f(k)\equiv \sin(k_x a)\sin(k_y a)$ (with $a$ the in-plane lattice constant) specifies the the fermion bilinear to which $\phi$ couples, in this case an anisotropic next-nearest-neighbor hopping. The precise form of $f(k)$ depends on microscopic details, but must respect the $B_{1g}$ symmetry of the nematic fluctuations. Since $f(k)$ vanishes at the high symmetry points $X,Y,M$, it is natural that the smaller pockets at X and Y have overall weaker coupling than the larger pocket at M.

We assume a mean-field form of $D(q)$, the static correlation function of $\phi$:
\eq{
D(q)=\frac{\chi_0}{1+\xi^2q^2}\Theta\left(\Lambda-|q|\right)
}
Here $\chi_0$ is the thermodynamic nematic susceptibility, $\xi$ is the correlation length, and we have introduced a hard momentum cutoff $\Lambda$ satisfying  $\xi^{-1}\ll\Lambda\ll k_1, k_2$. We will tune the strength of nematic fluctuations by tuning $\xi$ and $\chi_0$, which are related by $\chi_0\sim\xi^2$. Both $\xi$ and $\chi_0$ will increase as strontium concentration $x$ is decreased from $1.0$ to $0.7$ (i.e. as nematic correlations grow in strength). For explicitness we take $\chi_0$ to equal the (linearly-interpolated) nematic susceptibility measured by elastoresistance\cite{eckberg2019}.

Integrating out the boson, we obtain a four-fermion interaction  $V^{(ind)}_{k,p}=-\alpha^2f^2([k+p]/2) D(k-p)/4$. For $\Lambda\xi\gg 1,$ $D(q)$ is sharply peaked at $q=0$, and so can be approximated by a delta function of weight $W\equiv \int d^{d-1}q D(q)$, where the integral is over the Fermi surface. $W$ depends on the dimensionality of the nematic fluctuations, so we treat both the two and three dimensional cases\footnote{Even in a system whose electronic structure is highly two dimensional, nematic fluctuations may be rendered three dimensional by coupling to the lattice. However, the coupling to acoustic phonons introduces long range interactions that qualitatively alter the low energy physics\cite{Zacharias2015,Labat2017,Paul2017}, a complexity we ignore here for simplicity.}. 

Making these approximations, the full pairing interaction is $V_{k,p}=-V_0+V^{(ind)}_{k,p}$, or:
\eq{
V_{k,p}
\approx& -V_0-\frac{\alpha^2}{4} W f^2(k)\delta(k-p)\nn\\
\approx &-V_0\left(1+g f^2(k)\left[\frac{\delta(k-p)}{a}\right] \right),\label{eq:V_approx}\\
\text{where } g\equiv&\frac{\pi\chi_0\alpha^2a^{2-d}}{4V_0}\left(\frac{a}{\xi}\right)^{d-1}\label{eq:g}.
}

Here the dimensionless coupling constant $g$
sets the scale of fractional changes to the pairing eigenvalue. In $d=3$, $g$ also contains a factor $\log[1+(\Lambda\xi)^2]$.


We discretize the Fermi surface and numerically diagonalize the linearized gap equation, yielding the maximum pairing eigenvalue $\lambda$ and therefore $T_c$, both shown in Fig. \ref{fig:Tc}. The corresponding eigenfunction is the pair wave function, which is increasingly anisotropic as nematic fluctuations grow with decreasing $x$, as illustrated by Eq. \ref{eq:wavefunction}. This anisotropy can be experimentally identified by the splitting of peaks in the density of states, which occur at local maxima in the gap. The density of states for selected values of $x$ is shown in Fig. \ref{fig:dIdV} and can be measured by tunneling spectroscopy in either planar or scanning geometries. The gap anisotropy is also indirectly  measurable in the ratio of the zero temperature gap maximum to $T_c$, and in  the fractional specific heat jump at $T_c$, as shown in Fig. \ref{fig:ratio_jump}. The gap to $T_c$ ratio increases, while the specific heat jump decreases.\footnote{Though not obvious from the figure, the change in specific heat is parametrically smaller (in the weak coupling limit) than the other effects of the induced gap anisotropy, going as $\delta\lambda^2$ instead of $\delta\lambda$, as discussed in the supplementary material}

\emph{\bf Effects of B$_{1g}$ strain}: When nematic fluctuations are strong, the system is highly susceptible to the influence of a symmetry breaking field such as uniaxial strain. The effect of such strain on $T_c$ becomes increasingly dramatic at large $\chi_0$, but whether strain increases or decreases $T_c$ depends on microscopic details. Strain also has two qualitatively distinct effects on $T_c$: 1) It alters the band structure, leading to changes in $T_c$, as in Eq. \ref{eq:band}; and 2) It cuts off nematic fluctuations and alters the pairing interaction, as shown in Eqs. \ref{eq:cutoff} and \ref{eq:fluc}. 

Let strain couple to $\phi$ with coupling constant $\kappa$. At weak strain, $\phi$ acquires an expectation value $\langle\phi\rangle\approx \chi_0(\kappa\epsilon)$, with $\chi_0$ the thermodynamic nematic susceptibility as previously defined. Therefore, any physical quantity with B$_{1g}$ symmetry takes on a value proportional to $\epsilon$.
In particular the band structure acquires a (fractional) anisotropy of order $\alpha\chi_0\kappa\epsilon/E_F$. Such band structure effects alter any quantity that is nonzero at $\epsilon=0$, such as the density of states--and therefore $T_c$--by an amount proportional to the square of this anisotropy (linear variation is forbidden since such quantities are even under $\epsilon\to-\epsilon$)\footnote{The discussion in this section is largely general, but here we specialize to the case of a single-component order parameter. Multicomponent order parameters (for instance the degenerate $d_{xz}$ and $d_{yz}$ in a tetragonal system) permit a $T_c$ which varies as $|\epsilon|$.}. In particular, this effect changes $T_c$ by
\eq{
\frac{\left(\delta T_c\right)_{band}}{T_c} \propto\left(\frac{1}{\lambda_0}\right)\left(\frac{\alpha\kappa}{E_F}\right)^2\chi_0^2\epsilon^2
\label{eq:band}
}
where the proportionality constant is a dimensionless number of order one, whose magnitude and sign depend on microscopic details.

The application of strain alters not only the band structure, but the interactions. This is because strain cuts off nematic fluctuations, reducing the susceptibility $\chi_0$ and the correlation length $\xi$ by amounts proportional to $\epsilon^2$.
As the (zero-strain) $\chi_0$ increases, the coefficient of $\epsilon^2$ diverges as $\chi_0^{y_\epsilon}$, where $y_\epsilon\equiv 2+2\beta/\gamma=\{2.26\text{ in } d=2,\ 3\text{ in }d=3\}$ and $\beta,\gamma$ are the order parameter and susceptibility exponents, respectively. See the Supplementary Material for a derivation. Note that $y_\epsilon>2$, so these effects are parametrically stronger than the band structure effects described in Eq. \ref{eq:band}. Accordingly, the correlation length and susceptibility vary as
\eq{
\frac{\delta\chi_0}{\chi_0}\approx-(b_\chi u\kappa^2)\chi_0^{y_\epsilon}\epsilon^2,
\quad \frac{\delta\xi}{\xi}\approx-(b_\xi u\kappa^2)\chi_0^{y_\epsilon}\epsilon^2
\label{eq:cutoff}
}
Here $u>0$ is a dimensionful scale related to the self-interaction of $\phi$ (for instance the quartic term in Landau theory, as discussed in the Supplementary Material), and $b_{\chi,\xi}$ are dimensionless quantities. None of these quantities are specified by our model, but they are all positive since the symmetry breaking field $\epsilon$ cuts off nematic fluctuations.

Per \eqref{eq:V_approx}, the relative strength $g$ of the nematic-mediated interaction varies as the ratio of $\chi_0$ to $\xi^{d-1}$. Accordingly, $g$ (and therefore $T_c$) can either increase or decrease, since both $\chi_0$ and $\xi$ decrease with $\epsilon$. Up to logs,
\eq{
\frac{\left(\delta T_c\right)_{fluc}}{T_c} \propto+\left(\frac{[d-1]b_\xi-b_\chi}{\lambda_0}gu \kappa^2\right) \chi_0^{y_\epsilon}\epsilon^2
,
\label{eq:fluc}
}
so that this effect increases (decreases) $T_c$ for $[d-1]b_\xi >b_\chi$ ($[d-1]b_\xi<b_\chi$). We show in Fig. \ref{fig:strain} the second derivative of $T_c$ with respect to $\epsilon$, normalized by its value at $\epsilon=0$. Though the sign depends on microscopic details, the magnitude grows dramatically with decreasing $x$, much more so than the pairing eigenvalue (Fig. \ref{fig:Tc}) or the gap anisotropy (Figs. \ref{fig:dIdV}, \ref{fig:ratio_jump}).

\emph{\bf\bf Discussion}: We have shown that even weak coupling of the electrons near the Fermi surface to nematic fluctuations can explain the observed increase in the $T_c$ of \BNA \ as $x$ is reduced from 1.0 to 0.7. If this explanation is correct, there are a number of experimental consequences, as calculated above. In brief: 1) nematic fluctuations promote gap anisotropy, whose signatures are observable in tunneling conductance, specific heat, and the gap to $T_c$ ratio, among others; 2) uniaxial strain substantially affects $T_c$ and other superconducting properties, by altering both the band structure and the strength of the nematic-mediated interaction. The predictions regarding strain are asymptotically stronger effects than those regarding gap anisotropy, which could itself have explanations unrelated to nematic fluctuations. We therefore consider strain experiments to be the sharpest tests of the hypothesis of nematic-mediated pairing.

That said, it is intrinsically difficult to empirically establish a pairing mechanism based on the exchange of electronic fluctuations. In part, this is a matter of principle: Unlike the phonons of BCS theory, the electronic fluctuations considered are generally not well defined normal modes\footnote{Order parameter fluctuations \emph{are} well defined near a quantum critical point at which they order. This is essentially the scenario considered here, since we assume large $\chi_0$.}, so it isn't clear what it means for electrons to exchange them. But an even greater difficulty is that most forms of electronic fluctuations do not admit a simple tuning parameter to establish their relevance, as the isotope effect establishes the relevance of phonons. However, breaking the symmetry corresponding to a nearly ordered mode is essentially guaranteed to affect it much more than other aspects of the physics. As such, strain experiments to assess nematic-mediated pairing provide perhaps the most tightly controlled evaluations to date of (non-phonon) fluctuation-mediated pairing.

We close with a brief speculation regarding normal state properties. We have not computed these properties directly, but the symmetry of the nematic order parameter essentially guarantees substantial anisotropy\cite{Dellanna2006} of single particle properties measurable in photoemission, such as the effective mass and the linewidth. Such anisotropy is indeed found in numerical simulations of simple models for nematic quantum criticality\cite{lederer2017}. The strength of this anisotropy should show substantial dependence on the nematic fluctuation strength $\chi_0$. We eagerly await further investigations of the normal and superconducting state of this exciting material.
\label{discussion}

{\bf Acknowledgements} We gratefully acknowledge helpful discussions with Rafael Fernandes, Steve Kivelson, and Johnpierre Paglione, as well as data shared by JP and Christopher Eckberg. SL was supported in part by a Bethe/KIC fellowship at Cornell. EB was supported by the European Research Council (ERC) under grant HQMAT (grant no. 817799) and by the Israel-USA Binational Science Foundation (BSF). Theoretical studies by SL and E-AK were supported by the U.S. Department of Energy, Office of Basic Energy Sciences, Division of Materials Science and Engineering under Award DE-SC0018946.

\end{document}


\title{\bf Supplement to ``Tests of nematic-mediated superconductivity and the case of \BNA"}
\author{S. Lederer$^1$, Erez Berg$^2$, and Eun-Ah Kim$^1$}

\affiliation{$^1$Department of Physics, Cornell University, Ithaca, New York 14850, USA}
\affiliation{$^2$Department of Physics, Weizmann Institute of Science, Rehovot, Israel}
\maketitle

The main text computes quantities such as the gap to $T_c$ ratio and specific heat jump, as in Fig. 3. In Sec. \ref{sec:sup_anisotropic} we derive the formulas for these quantities for an anisotropic (i.e. momentum-dependent) gap function $f(k)$, which reduce to the familar BCS values when $f(k)=1$.  In Sec. \ref{sec:sup_scaling}, we show how uniaxial strain cuts off nematic fluctuations using scaling arguments. In Sec. \ref{sec:sup_strain} we compute the various terms that alter $T_c$ at small strain including the effects of retardation.

\section{$2\Delta_0/T_c$ and $\Delta C/C$ for an anisotropic gap function}
\label{sec:sup_anisotropic}
Our basic tool will be the BCS gap equation for a singlet superconductor:
\eq{
\Delta_k =\int d^dp V_{k,p}\frac{\Delta_p}{2E_p}\tanh\left[\frac{E_p}{2T}\right]
}
where $V$ is $-1$ times the (attractive) pairing interaction in the Cooper channel, assumed to be weak, and $E_p \equiv \sqrt{\epsilon_p^2+|\Delta_p|^2}$ with $\epsilon_p$ the single particle dispersion. We assume that $V_{k,p}$ is zero unless $|\epsilon_k|,|\epsilon_p|<\Omega$, where $\Omega$ is some cutoff much smaller than UV scales such as $E_F$, and is the analog of the Debye frequency. We also assume that, besides this cutoff, there is no dependence of $V_{k,p}$ on the components of $k$ and $p$ perpendicular to the Fermi surface. 
\subsection{Critical temperature $T_c$}
At $T_c$, $\Delta_k=0$ and we can linearize the gap equation to determine $T_c$ as well as the pair wavefunction:
\eq{
\Delta_k =\int d^dp V_{k,p}\frac{\Delta_p}{2\epsilon_p}\tanh\left[\frac{\epsilon_p}{2T_c}\right]
}
We first integrate over $p_\perp$, the component of momentum perpendicular to the Fermi surface:
\eq{
\Delta_k =&\int d^{d-1}p_\parallel V_{k,p}\Delta_p\int \frac{dp_\perp} {2\epsilon_p}\tanh\left[\frac{\epsilon_p}{2T_c}\right]\nn\\
=&\int \frac{d^{d-1}p_\parallel}{v_F(p)} V_{k,p}\Delta_p\int_0^{\Omega} \frac{d\epsilon} {\epsilon}\tanh\left[\frac{\epsilon}{2T_c}\right]\nn\\
=&\int \frac{d^{d-1}p_\parallel}{v_F(p)} V_{k,p}\Delta_p\int_0^{\Omega/T_c} \frac{dx} {x}\tanh\left[\frac{x}{2}\right]\nn\\
\approx &\log\left[\frac{1.134 \Omega}{T_c}\right] \int \frac{d^{d-1}p_\parallel}{v_F(p)} V_{k,p}\Delta_p,
}
Where by $v_F(p)$ we mean the norm of the Fermi velocity, and in the final line the approximation becomes exact in the limit $\Omega/T_c\to \infty$. We now rewrite the above, defining the pair wavefunction $\phi_k\propto\Delta_k/\sqrt{v_F(k)}$
\eq{
\phi_k =&\log\left[\frac{1.134 \Omega}{T_c}\right]\int d^{d-1}p_\parallel \Gamma_{k,p}\phi_p\text{, where}\\
\Gamma_{k,p}\equiv & \frac{V_{k,p}}{\sqrt{v_F(k)v_F(p)}}
}
$\Gamma_{k,p}$ is a Hermitian matrix, and therefore has real eigenvalues. Now let $\phi_k$ be the eigenfunction of $\Gamma$ with largest eigenvalue $\lambda$:
\eq{
\phi_k \approx &\log\left[\frac{1.134 \Omega}{T_c}\right]\lambda \phi_k\text{, so}\nn\\
T_c\approx &1.134\Omega\times \exp\left[-\frac{1}{\lambda}\right]
}
Unless $\Gamma_{k,p}$ has unusual structure, the splitting between its largest eigenvalue $\lambda$ and smaller ones will be of order $\lambda$. Therefore the effective $T_c$ of other channels will be exponentially smaller than the $T_c$ of the optimal channel, and the gap function at all temperatures will be determined by $\phi_k$. From now on, we will assume that $\phi_k$ is normalized, and that the other pairing channels of $\Gamma_{k,p}$ can be neglected:
\eq{
\int d^{d-1}p_\parallel |\phi_p|^2= & 1\nn\\
 \Gamma_{k,p}\approx & \lambda \phi_k\phi_p^*\label{eq:gamma}
}
\subsection{Zero temperature gap function}
At temperature $T$, the gap function will be given by $\Delta_{0} (T) f(k)$, for some real $\Delta_{0}(T)$ and a dimensionless function $f(k)$ determined by the pair wavefunction. Explicitly, 
\eq{
f(k)&= \phi_k \sqrt{v_F(k)}/\alpha\nn\\
\phi_k &=\alpha \frac{f(k)}{\sqrt{v_F(k)}}\nn\\
\int d^{d-1}k_\parallel |\phi_k|^2 = 1 &= \alpha^2 \int \frac{d^{d-1}k_\parallel}{v_F(k)}|f(k)|^2\nn\\
\alpha^2 &=\left( \int \frac{d^{d-1}k_\parallel}{v_F(k)}|f(k)|^2\right)^{-1}
\label{eq:alpha}
}
At zero temperature, the $\tanh$ in the gap equation is equal to $1$ and we can write:
\eq{
\Delta_0 f(k)  =\Delta_0  \int d^dp V_{k,p}\frac{f(p)}{2E_p}
}
As before, we proceed to integrate over $p_\perp$:
\eq{
 f(k)  = &\int \frac{d^{d-1}p_\parallel}{v_F(p)} V_{k,p}f(p) \int_0^\Omega d\epsilon \frac{1}{\sqrt{\epsilon^2+|\Delta_p|^2}}\nn\\
=& \int \frac{d^{d-1}p_\parallel}{v_F(p)} V_{k,p}f(p) \int_0^{\Omega/|\Delta_p|}  \frac{dx}{\sqrt{x^2+1}}\nn\\
=& \int \frac{d^{d-1}p_\parallel}{v_F(p)} V_{k,p}f(p)\sinh^{-1}\left[\frac{\Omega}{|\Delta_p|}\right]\nn\\
\approx& \int \frac{d^{d-1}p_\parallel}{v_F(p)} \Gamma_{k,p}\sqrt{v_F(k)v_F(p)}f(p)\left(\log\left[\frac{2\Omega}{\Delta_0}\right]-\log\left[ |f(p)|\right]\right),
}
Where the approximation in the final line is exact in the weak coupling limit $\Omega/\Delta_0\to\infty$, and we have used the definition of $\Gamma$ for convenience. Some rewriting:
\eq{
\frac{f(k)}{\sqrt{v_F(k)}}\approx& \int d^{d-1}p_\parallel \Gamma_{k,p}\frac {f(p)}{\sqrt{v_F(p)}}\left(\log\left[\frac{2\Omega}{\Delta_0}\right]-\log\left[ |f(p)|\right]\right)
}
Now rewrite $\Gamma$ using Eq. \ref{eq:gamma}:
\eq{
\Gamma_{k,p}\approx \lambda \phi_k\phi^*_p=\lambda\frac{\alpha f(k)}{\sqrt{v_F(k)}} \cdot  \frac{\alpha f^*(p)}{\sqrt{v_F(p)}},
\label{eq:gamma2}
}
And substitute this equation into the previous:
\eq{
\frac{f(k)}{\sqrt{v_F(k)}}\approx& \lambda\alpha^2\int d^{d-1}p_\parallel\frac {f(k)}{\sqrt{v_F(k)}}\frac {f^*(p)}{\sqrt{v_F(p)}}\frac {f(p)}{\sqrt{v_F(p)}}\left(\log\left[\frac{2\Omega}{\Delta_0}\right]-\log\left[ |f(p)|\right]\right)
}
Now we can divide the equation by the quantity on the left, and use the expression for $\alpha$ of Eq. \ref{eq:alpha}:
\eq{
1\approx& \lambda\left( \int \frac{d^{d-1}k_\parallel}{v_F(k)}|f(k)|^2\right)^{-1}\int\frac{ d^{d-1}p_\parallel}{v_F(p)}|f(p)|^2\left(\log\left[\frac{2\Omega}{\Delta_0}\right]-\log |f(p)|\right)\nn\\
=&\lambda\left(\log\left[\frac{2\Omega}{\Delta_0}\right]-\frac{\langle |f |^2 \log |f |\rangle_{FS}
}{\langle|f|^2\rangle_{FS} }\right)}
Where the angle brackets indicate a Fermi surface average weighted by the density of states:
\eq{
\langle {h} \rangle_{FS}\equiv \frac{\int \frac{d^{d-1}p_\parallel}{v_F(p)}h(p)}{\int \frac{d^{d-1}p_\parallel}{v_F(p)}}}
 Now we can solve for $\Delta_0$:
\eq{
\Delta_0=2\Omega\exp{\left[-\frac{1}{\lambda}-\frac{\langle|f|^2\log|f|\rangle_{FS}}{\langle|f|^2\rangle_{FS}}\right]}
}
As expected $\Delta_0 \propto T_c$, and
\eq{
\frac{2\Delta_0}{T_c}\approx 3.527 \exp\left[-\frac{\langle|f|^2\log|f|\rangle_{FS}}{\langle|f|^2\rangle_{FS}}\right]
}
For the case $f(k)=1$, the bracketed exponent is zero, and we recover the standard BCS result. 

\subsection{Temperature dependence of gap near $T_c$}
Near $T_c$, the gap magnitude has a mean-field temperature dependence proportional to $\sqrt{T_c-T}$. We now derive the prefactor, which, like $2\Delta_0/T_c$, will depend on the momentum space structure of $\Delta_k\equiv \Delta_0 f(k)$.

\eq{
\Delta_k =&\int d^dp V_{k,p}\frac{\Delta_p}{2E_p}\tanh\left[\frac{E_p}{2T}\right]\nn\\
 f(k) = &\int d^dp V_{k,p}\frac{f(p)}{2E_p}\tanh\left[\frac{E_p}{2T}\right]
}
We now differentiate both sides with respect to temperature. It is convenient to introduce $x\equiv E_p/(2T)$.
\eq{
0 = &\int d^dp V_{k,p}f(p)\left[ \partial_{E_p}\left(\frac{1}{2E_p} \tanh\left[\frac{E_p}{2T}\right]\right)(\partial_T E_p) +\frac{1}{2E_p}\partial_x \tanh(x)(\partial_T x ) \right]
}
Now we rearrange, and use $\partial_T x = -E_p/(2T^2)$,  $\partial_T E_p= (\partial_T E_p^2)/(2E_p) = (\partial_T |\Delta_p|^2)/(2E_p)$
\eq{
\int d^dp V_{k,p}f(p)\left[\frac{1}{4T^2}\partial_x \tanh(x)\right] = &\int d^dp V_{k,p}f(p)\left[ \frac{1}{2E_p} \partial_{E_p}\left(\frac{1}{2E_p} \tanh\left[\frac{E_p}{2T}\right]\right)\right]\partial_T|\Delta_p|^2\nn\\
 = &\partial_T (\Delta_0^2)\int d^dp V_{k,p}f(p)|f(p)|^2\left[ \frac{1}{2E_p} \partial_{E_p}\left(\frac{1}{2E_p} \tanh\left[\frac{E_p}{2T}\right]\right)\right]
\label{eq:dDelta}}
We can now evaluate $\partial_T (\Delta_0^2)$ at $T_c$ by taking the ratio of the left to the right integrals. At $T_c$, $E_p$ becomes $|\epsilon_p|$ and we can do the $p_\perp$ integrals. First on the left side:
\eq{
\int dp_\perp \left[\frac{1}{4T^2}\partial_x \tanh(x)\right] =\frac{1}{2Tv_F}\int dx \partial_x \tanh(x)=\frac{1}{Tv_F} \tanh\left[\frac{\Omega}{2T}\right] \approx \frac{1}{Tv_F}
}
For the right side:
\eq{
\int dp_\perp \left[ \frac{1}{2E_p} \partial_{E_p}\left(\frac{1}{2E_p} \tanh\left[\frac{E_p}{2T}\right]\right)\right]=&\frac{1}{v_F(p)}\int d\epsilon\left[ \frac{1}{2\epsilon} \partial_{\epsilon}\left(\frac{1}{2\epsilon} \tanh\left[\frac{\epsilon}{2T}\right]\right)\right]\nn\\
=&\frac{1}{16 T^2 v_F(p)}\int\frac{dx}{x}\partial_x \left(\frac{\tanh(x)}{x}\right)\nn\\
=&\frac{-0.107}{T^2v_F(p)},
}
Where the integral has been computed numerically. Putting these results back into equation \eqref{eq:dDelta} yields:
\eq{
\int \frac{d^{d-1}p_\parallel}{v_F(p)} V_{k,p}f(p)\left[\frac{1}{T}\right] =-\partial_T (\Delta_0^2) \int \frac{d^{d-1}p_\parallel}{v_F(p)} V_{k,p}f(p)|f(p)|^2\left[\frac{0.107}{T^2}\right]
}
We now rewrite $V_{k,p}$ using Eqs. \eqref{eq:gamma} and \eqref{eq:gamma2} and take the quotient:
\eq{
\frac{1}{T_c}\partial_T(\Delta_0)^2 =& -9.384 \left(\int \frac{d^{d-1}p_\parallel}{v_F(p)}|f(p)|^2\right)\left(\int \frac{d^{d-1}p_\parallel}{v_F(p)} |f(p)|^4\right)^{-1}\nn\\
=&-9.384\left(\frac{\langle |f|^2\rangle_{FS}}{\langle |f|^4\rangle_{FS}}\right),
\label{eq:deriv}}
\subsection{Specific heat jump}
First we write entropy density for a gas of spin-degenerate Bogoliubov quasiparticles with dispersion $E_k$. At a given momentum $k$, there is a probability $f(E_k/T)$ that the state is occupied and $1-f(E_k/T)=f(-E_k/T)$ that the state is unoccupied, where $f(x)=(1+\exp(x))^{-1}$ is the Fermi distribution function. Accordingly the entropy density is
\eq{
S=-2 \int d^dk \left(f(E_k) \log[f(E_k)] + f(-E_k)\log[f(-E_k)]\right)
}
The specific heat $C=T\partial_T S$, so we differentiate the above with respect to T introducing $x=E_k/T$:
\eq{
\partial_T S=&-2 \int d^dk\partial_x \left(f(x) \log[f(x)] + f(-x)\log[f(-x)]\right)\partial_T x\nn\\
=&-2 \int d^dk \left(\partial_x f(x)\right)\log\left[\frac{f(x)}{f(-x)}\right]\partial_T x\nn\\
=&-2 \int d^dk \left(-\frac{e^x}{(1+e^x)^2}\right)(-x)\partial_T x
}

Now use $\partial_T x=\partial_T E_k/T-x/T$, with $\partial_T E_k= (\partial_T E_k^2)/(2E_k) = (\partial_T |\Delta_k|^2)/(2Tx)$:
\eq{
\partial_T S=&\frac{2}{T} \int d^dk \left(\frac{e^x}{(1+e^x)^2}\right)\left(x^2-\frac{\partial_T|\Delta_k|^2}{2T}\right)\nn\\
=&\frac{1}{2T} \int d^dk \cosh^{-2}\left[\frac{x}{2}\right] \left(x^2-\frac{\partial_T|\Delta_k|^2}{2T}\right)
}
We now evaluate this just below $T_c$, where $E_k=|\epsilon_k|$, performing the $k_\perp$ integrals first:
\eq{
\partial_T S
=&\frac{1}{2} \int \frac{d^{d-1}k_\parallel}{v_F(k)} \cdot  \int dx \cosh^{-2}\left[\frac{x}{2}\right] \left(x^2-\frac{\partial_T|\Delta_k|^2}{2T}\right)\nn\\
=& \int \frac{d^{d-1}k_\parallel}{v_F(k)}\left\{ \frac{2\pi^2}{3}-\frac{\partial_T|\Delta_k|^2}{T} \right\}
}
We now substitute \eqref{eq:deriv}:
\eq{
\partial_T S=& \int \frac{d^{d-1}k_\parallel}{v_F(k)}\left\{ \frac{2\pi^2}{3}+  9.384 |f(k)|^2 \left(\frac{\langle |f|^2\rangle_{FS}}{\langle |f|^4\rangle_{FS}}\right)\right\}\nn\\
\frac{C}{T}=\partial_T S=&\frac{2\pi^2}{3}\rho(E_F) \left\{1+1.425  \left(\frac{\langle |f|^2\rangle_{FS}^2}{\langle |f|^4\rangle_{FS}}\right) \right\}
}
The first term in brackets is the same just above and just below $T_c$, whereas the second term is only present in the superconducting state. The fractional jump in specific heat is therefore
\eq{
\frac{\Delta C}{C}=1.425\left(\frac{\langle |f|^2\rangle_{FS}^2}{\langle |f|^4\rangle_{FS}}\right),
}
Where $1.425$ is, as expected, the BCS result. Generally, anisotropy will reduce this value. Suppose the gap is weakly anisotropic, with $f(k)=1+\delta(k)$ for some small, real $\delta(k)$. Then

\eq{
\frac{\Delta C}{C}=&\frac{1.425}{\rho(E_F)}\left(\int \frac{dk_\parallel}{v_F}[1+\delta(k)]^2  \right)^2\left(\int \frac{dk_\parallel}{v_F}[1+\delta(k)]^4\right)^{-1}\nn\\
=&\frac{1.425}{\rho(E_F)}\left(\int \frac{dk_\parallel}{v_F}[1+2\delta(k)+\dots]  \right)^2\left(\int \frac{dk_\parallel}{v_F}[1+4\delta(k)+\dots]\right)^{-1}\nn\\
=&\frac{1.425}{\rho(E_F)}\left(\rho(E_F)+2\int \frac{dk_\parallel}{v_F}\delta(k)+\dots]  \right)^2\left(\rho(E_F)+4\int \frac{dk_\parallel}{v_F}\delta(k)+\dots] \right)^{-1}\nn\\
=&1.425+{\cal O}\left(\delta^2\right),
}
where the dots in the middle lines mean terms of order $\delta^2$. So the deviation of the specific heat jump from its BCS value is proportional to the square of the anisotropy. For the situation discussed in the paper, this means the change in specific heat is of order $\delta\lambda^2$, and therefore parametrically smaller than the other effects discussed.
\section{Scaling of bosonic observables with symmetry breaking field}
\label{sec:sup_scaling}
Let $\phi$ be the bosonic field in question, $r$ the tuning parameter, and $h$ the symmetry breaking field. We are ultimately interested in $W$, the integral of the bosonic correlator over the $d-1$ dimensional Fermi surface, i.e. the weight of the delta function that approximates the forward scattering interaction mediated by $\phi$. This will be proportional to the thermodynamic susceptibility $\chi_0$ divided by some increasing function of the correlation length $\xi$. At nonzero $r$ and for small $h$, $W$ must depend quadratically on $h$, and we are interested in the scaling of the coefficient of that quadratic dependence with $r$. We begin with mean field theory

\subsection{Mean field theory}
Start with the Ginzburg-Landau free energy
\eq{
f=\frac{\kappa}{2}(\nabla\phi)^2+\frac{r}{2}\phi^2+\frac{u}{4}\phi^4-h\phi
}
Taking a spatially uniforn solution and minimizing the free energy with respect to $\phi$ yields
\eq{
r\phi+u\phi^3=h
}
In the small $h$ limit, and for $r>0$, it follows that $\phi=\chi_0(r,0) h+{\cal O}(h^3)$, with $\chi_0(r,0)=1/r$. Now implicitly differentiate with respect to $h$, with $\partial_h \phi = \chi_0$:
\eq{
(r+3u\phi^2)\chi_0(r,h)=&1\nn\\
\Rightarrow \chi_0(r,h)=&\frac{1}{r+3u\phi^2}\approx \frac{1}{r}\left(1-\frac{3u\phi^2}{r}\right)\nn\\
=&\chi_0(r,0)\left(1-\frac{3u}{r}[\chi_0(r,0)h]^2\right)\nn\\
=&\chi_0(r,0)\left(1-3u\chi_0(r,0)^3h^2\right)
}
So the fractional change in susceptility goes like $h^2\cdot\chi^3$. Restoring the gradients gives the same result for the correlation length. This a limiting  case of the general scaling theory discussed below.

\subsection{Scaling of symmetry breaking field with tuning parameter}
Suppose we are in a spatial dimension for which the ``bare" critical theory of the nematic boson obeys scaling. Then the magnetization has the form
\eq{
M(r,h)\sim r^{\beta}m(hr^{x_h}),
}
where $\beta$ is the order parameter critical exponent, $m(u)$ is a universal function which goes to a nonzero value as $u\to 0$, and $x_h$ is an exponent we will now relate to the standard critical exponents. The susceptibility is
\eq{
\chi=&\frac{\partial}{\partial h} M(r,h)= r^{\beta+x_h} m'(hr^{x_h})\nn\\
\sim&r^{-\gamma}\nn\\
\Rightarrow&x_h=-(\gamma+\beta)
}
\subsection{Scaling of weight $W$}
Form the static correlation function of $\phi$ in momentum space, which will satisfy a multi-variable scaling function:
\eq{
G(q,r,h)\equiv \int d\tau \int dx e^{-q x} \langle\phi(x,\tau)\phi(0,0)\rangle\sim r^{-\gamma} g[q\xi_r,hr^{x_h}]
}
Here $\gamma$ is the susceptibility critical exponent, $\xi_r\sim r^{-\nu}$ is the $h=0$ correlation length, $x_h$ is as in the previous section, and $g[u,v]$ is a universal function, with $g[0,v]\ (g[u,0])$ an analytic function of $v\ (u)$. Since $G(q,r,h)$ is a scaling function of two variables, there are actually two different length scales which diverge in the $h,r\to 0$ limit, $\xi_r\sim r^{-\nu}$ and $\xi_h\sim h^{-\nu/(\gamma+\beta)}$. At finite $r$, and in the $h\to 0$ limit, the relevant length scale is $\xi_r$, whereas the relevant length scale is $\xi_h$ in the $r\to 0$ limit at finite $h$.

For purposes of understanding the electronic forward scattering interaction mediated by $\phi$, we are interested in $G(q,r,h)$ integrated over the $d-1$ dimensions transverse to the Fermi surface:
\eq{
W\equiv&\int d^{d-1} q G(q,r,h)\nn\\
\sim&r^{-\gamma}\int [d^{d-1} q] g(q\xi_r,hr^{x_h})\nn\\
=&\frac{r^{-\gamma}}{\xi_r^{d-1}}\int [d^{d-1} (q\xi_r)] g(q\xi_r,hr^{x_h})\nn\\
\equiv&r^{-\gamma+\nu(d-1)}w(hr^{x_h})
}
where in the final line we have defined a scaling function $w$, assuming the integral is convergent. This is true in $d=2$, and marginally false in $d=3$, so that the function $w$ also depends logarithmically on $\Lambda\xi$, with $\Lambda$ is a momentum cutoff. In either case, $W$ must depend quadratically on $h$ for finite $r$ and in the limit $h\to 0$:
\eq{
W(r,h)\approx & W(r,0)\left[1+a_W (hr^{x_h})^2\right]\nn\\
=&W(r,0)\left[1+a_W \left(hr^{-(\gamma+\beta)}\right)^2\right]\nn\\
=&W(r,0)\left[1+a_W h^2\chi^{2(\gamma+\beta)/\gamma}\right]
,}
where $a_W$ is a nonuniversal coefficient. For the nematic case, $h\propto \epsilon$, where $\epsilon$ is the appropriate uniaxial strain, so that $y_\epsilon=2x_h/\gamma=2+\beta/\gamma$. If the propagator is a free boson with susceptibility $\chi_0$ and correlation length $\xi$, $W\propto \chi_0 \xi^{1-d}$ and therefore 
\eq{
\frac{W(r,h)-W(r,0)}{W(r,0)}\propto \frac{d\chi_0}{\chi_0}-(d-1)\frac{d\xi}{\xi},
}
corresponding to Eq. 11 of the main text.

\section{Strain effects on $T_c$}
\label{sec:sup_strain}
Here, we compute the leading dependence of $T_c$ on strain, including the effects of retardation. Let the energy scales of the ``bare" and nematic-mediated interaction be  $\Omega_0$ and $\Omega$, respectively. $T_c$ is given by

\eq{
T_c=&\Omega\exp\left[-\frac{1}{\lambda^*}\right]\text{, with}\\
\lambda^*=\lambda_0^*+\delta\lambda,\quad &\lambda_0^*=\frac{\lambda_0}{1-\lambda_0\log[\Omega_0/\Omega]}}

With the eigenvalue shift
\eq{
\delta\lambda\approx & \lambda_0\times gF \times  \begin{cases}\frac{\xi}{a},\quad&\text{d=2}\\\\
\log[1+(\Lambda\xi)^2],&\text{d=3}\end{cases}\label{deltalambda}\\\\
 F\equiv &\left(\frac{1}{a}\right)\left(\int_{FS} \frac{dk}{v_F}\frac{f^2(k)}{v_F}\right)\left(\int_{FS} \frac{dk}{v_F}\right)^{-2},\quad g\equiv
\frac{\pi\alpha^2\chi_0a^{4-d}}{4V_0\xi^2}
}
The fractional change in $T_c$ is
\eq{
\frac{dT_c}{T_c}=&\frac{d\Omega}{\Omega}+\frac{d\lambda^*}{(\lambda^*)^2}+\dots,
}
Where $\dots$ represents higher order terms. We will compute the above in terms of several microscopic parameters, some of them of known sign:
\eq{
\frac{d\xi}{\xi}=-\frac{1}{z}\frac{d\Omega}{\Omega}=-b_\xi(u\kappa^2)\chi_0^{2(\gamma+\beta)/\gamma}\epsilon^2,&\quad \frac{d\chi_0}{\chi_0}=-b_{\chi}(u\kappa^2)\chi_0^{2(\gamma+\beta)/\gamma}\epsilon^2 \label{fluc_strain}\\
\frac{d\lambda_0}{\lambda_0}=a_\lambda\left(\frac{\alpha\kappa}{E_F}\right)^2\chi_0^2\epsilon^2,&\quad
\frac{d F}{ F}=a_F\left(\frac{\alpha\kappa}{E_F}\right)^2\chi_0^2\epsilon^2\label{op_strain}
}
In the upper line, $z$ is the dynamical critical exponent of the proximate nematic QCP ($z=1$ when this QCP is governed by the transverse-field Ising fixed point). For the upper line, the sign is determined, since strain cuts off the nematic fluctuations; the magnitude is unknown. In the lower line neither sign nor magnitude are known without a microscopic model. 

\subsection{Change in logs}
\eq{
d\log\left[\frac{\Omega_0}{\Omega}\right]=&\log\left[\frac{\Omega_0}{\Omega(1+d\Omega/\Omega)}\right]-\log\left[\frac{\Omega_0}{\Omega}\right]\nn\\
\approx&-\frac{d\Omega}{\Omega}
}
Similarly
\eq{
d\log\left[1+(\Lambda\xi)^2\right]\approx\frac{2d\xi}{\xi}
}
\subsection{Change in $\lambda_0^*$}
Now we're in a position to compute the change in $\lambda_0^*$:
\eq{
d\lambda_0^*=&\frac{\partial \lambda_0^*}{\partial\Omega}d\Omega+\frac{\partial \lambda_0^*}{\partial\lambda_0}d\lambda_0}
Straightforward differentiation yields
\eq{
\frac{\partial \lambda_0^*}{\partial\Omega}=&
-\frac{(\lambda_0^*)^2}{\Omega}\\
\frac{\partial \lambda_0^*}{\partial\lambda_0}
=&\left(\frac{\lambda_0^*}{\lambda_0}\right)^2
}

All told:
\eq{
d\lambda_0^*=&-(\lambda_0^*)^2\left(\frac{d\Omega}{\Omega}-\frac{1}{\lambda_0}\frac{d\lambda_0}{\lambda_0}\right)
}

\subsection{Change in $\delta\lambda$}
Start with Eq. \ref{deltalambda}:
\eq{
d\delta\lambda=&\delta\lambda\times\begin{cases}\frac{d\chi_0}{\chi_0}+\frac{dF}{F}-\frac{d\xi}{\xi},\quad&d=2\\ \\
\frac{d\chi_0}{\chi_0}+\frac{dF}{F}-2\frac{d\xi}{\xi}+\frac{d\log[1+(\Lambda\xi)^2]}{\log[1+(\Lambda\xi)^2]},&d=3\end{cases}\nn\\
=&\delta\lambda\times\begin{cases}\frac{d\chi_0}{\chi_0}+\frac{dF}{F}-\frac{d\xi}{\xi},\quad&d=2\\ \\
\frac{d\chi_0}{\chi_0}+\frac{dF}{F}-2\frac{d\xi}{\xi}+\frac{d\xi}{\xi\log[\Lambda\xi]},&d=3\end{cases}
}
\subsection{Net change}
Now for the full eigenvalue:
\eq{
d\lambda^*=&d\lambda_0^*+d\delta\lambda\nn\\
=&-(\lambda_0^*)^2\left(\frac{d\Omega}{\Omega}-\frac{1}{\lambda_0}\frac{d\lambda_0}{\lambda_0}\right)+\delta\lambda\times\begin{cases}\frac{d\chi_0}{\chi_0}+\frac{dF}{F}-\frac{d\xi}{\xi},\quad&d=2\\ \\
\frac{d\chi_0}{\chi_0}+\frac{dF}{F}-2\frac{d\xi}{\xi}+\frac{d\xi}{\xi\log[\Lambda\xi]},&d=3\end{cases}
}

\eq{
\frac{dT_c}{T_c}=&\frac{d\Omega}{\Omega}+\frac{d\lambda^*}{(\lambda^*)^2}\nn\\
=&\frac{d\Omega}{\Omega}\left[1-\left(\frac{\lambda_0^*}{\lambda^*}\right)^2\right]+\left(\frac{\lambda_0^*}{\lambda^*}\right)^2\frac{1}{\lambda_0}\frac{d\lambda_0}{\lambda_0}+
\frac{\delta\lambda}{(\lambda^*)^2}\times\begin{cases}\frac{d\chi_0}{\chi_0}+\frac{dF}{F}-\frac{d\xi}{\xi},\quad&d=2\\ \\
\frac{d\chi_0}{\chi_0}+\frac{dF}{F}-2\frac{d\xi}{\xi}+\frac{d\xi}{\xi\log[\Lambda\xi]},&d=3\end{cases}
}
There are a number of small parameters in this expression, and therefore several parametrically broad regimes in which various terms above might be important. Keeping only the parametrically largest terms and substituting the expressions of Eqs. \ref{fluc_strain} and \ref{op_strain} yields
\eq{
\frac{dT_c}{T_c}=&\frac{a_\lambda}{\lambda_0}\left(\frac{\alpha\kappa}{E_F}\right)^2\chi_0^2\epsilon^2+
\frac{\delta\lambda}{(\lambda_0^*)^2}\times\bigg((d-1)b_\xi-b_{\chi}\bigg)(u\kappa^2)\chi_0^{2(\gamma+\beta)/\gamma}\epsilon^2\nn\\
=&\left\{\frac{a_\lambda}{\lambda_0}\left(\frac{\alpha\kappa}{E_F}\right)^2\chi_0^2+
\frac{\delta\lambda}{(\lambda_0^*)^2}\times\bigg((d-1)b_\xi-b_{\chi}\bigg)(u\kappa^2)\chi_0^{2(\gamma+\beta)/\gamma}\right\}\epsilon^2
}
The first term is the contribution from the induced nematic order parameter, and the second from the cutting off of nematic fluctuations. The latter comes with a larger power of $\chi_0$ than the first, but with a coefficient that may be parametrically smaller. Note that terms proportional to the change in the frequency scale $\Omega$ are negligible compared to terms involving changes in the pairing eigenvalue.